\documentstyle[12pt, cite, epsfig, a4]{article}

\newcommand{\la}{\langle}
\newcommand{\ra}{\rangle}

\newcommand{\beq}{\begin{equation}}
\newcommand{\eeq}{\end{equation}  }

\begin{document}

\hfill Nijmegen preprint

\hfill HEN-392

\hfill April 1996

\vspace{1cm}
\begin{center}

{\Large\bf 
Bin-Bin Correlation Measurement by the Bunching-Parameter Method}
\vspace{2.cm}

{\large S.V.Chekanov\footnote[1]{On leave from
Institute of Physics,  AS of Belarus,
Skaryna av.70, Minsk 220072, Belarus}, W.Kittel}

\medskip

{\it High Energy Physics Institute Nijmegen
(HEFIN), University of Nijmegen/NIKHEF,\\
NL-6525 ED Nijmegen, The Netherlands}

\bigskip

{\large V.I.Kuvshinov}

\medskip

{\it Institute of Physics,  AS of Belarus,
Skaryna av.70, Minsk 220072, Belarus}

\vspace{1.0cm}

{\large\em To appear in Acta Phys. Pol.}

\vspace{1.0cm}

\end{center}

\begin{abstract}
A new method for the 
experimental study of bin-bin correlations is proposed.
It is shown that this method is able to reveal important additional
information on bin-bin correlations, 
beyond that of factorial-correlator 
measurements.
\end{abstract}

\section{Introduction}
\label{sec:abi}

In order to obtain a comprehensive knowledge of the dynamics of 
particle production  
in high-energy reactions, two aspects of
multiplicity fluctuations need to be studied: 

1) the dependence of the multiplicity distribution (or its
characteristics) on the size of the phase-space interval;

2) the dynamical correlations between two or more bins where
this dependence is investigated.

The first point corresponds to the measurement of
the local fluctuations, the second one to a simultaneous measurement of 
the local characteristics in two (or more) different bins in order to
reveal correlations between these local fluctuations. If no 
correlations exist between  fluctuations in different bins, then 
complete information on an experimental sample can be obtained from
local fluctuation measurements.

Dynamical information on fluctuations 
in a system with an infinite
number of particles per event can be obtained from the multivariate
density probability distribution $P(\rho_1, \rho_2, ..\rho_M)$,
where $\rho_m$ is the particle density in bin $m$ ($m=1,\ldots ,M$).
This distribution can be studied by constructing the multivariate
moments $\la \rho_1^{q_1} \rho_2^{q_2}\ldots \rho_M^{q_M} \ra$.
Due to the very complex structure of this quantity, however, 
one usually resorts to the study of only two moments: $\la \rho_m^{q}\ra$
and $\la \rho_m^{q} \rho_{m'}^{q'}\ra$, which contain a small
fraction of the information on dynamical fluctuations in a system. 
The bivariate moment 
$\la \rho_m^{q_m} \rho_{m'}^{q'}\ra$
contains the information on bin-bin correlations.

In practice, bin-bin correlations always exist, i.e., 
$\la \rho_m^{q} \rho_{m'}^{q'}\ra \ne 
\la \rho_m^{q}\ra \la \rho_{m'}^{q'}\ra$,  since final-state particles
are not produced  independently of each other.  
The production of a particle at high energy usually enhances the 
probability of producing other particles. 
The number of particles observed in a given phase-space
bin, therefore, is always affected  by the number  of
particles found in other bins.
Moreover, there are  more trivial 
(statistical) reasons for the observation of correlations in a system
of finite fixed final-state multiplicity: for such a system,
finding a particle in a single bin is less probable if 
another particle has already been counted in another bin.
The latter case has no dynamical reason, but can influence the
correlations observed in such a system. 

In \cite{abir1},  
Bia\l as  and Peschanski have adapted the method of normalized 
factorial moments to the measurement of 
dynamical bin-bin correlations by means of
factorial correlators. The use of these quantities, 
as well as of the normalized
factorial moments,  has  mainly been motivated by the
Poissonian-noise suppression \cite{abir2}, thereby  
opening the possibility of modeling intermittency
phenomena and bin-bin correlations by means of continuous densities. 

In this paper we propose  another experimental tool to measure  bin-bin 
correlations  by means of the 
bunching-parameter approach \cite{abir3,abir4,abir5,abir6}. 
In the following,
we shall discuss the experimental advantages of using such a method
(Sect.~2). As an  illustration, the bin-bin correlation measurement
by the lowest-order bunching correlator is given in Sect.~3.

\section{Bunching correlators}

One of the characteristic 
features of any local multiplicity fluctuations in 
high-energy physics is the existence of bin-bin correlations.
If we have two non-overlapping bins, $m$ and $m'$ of size
$\delta$, then the discrete two-dimensional multiplicity distribution
$P_{n,n'}^{m,m'}(\delta )$ having
$n$ and $n'$ particles in bins $m$ and $m'$, respectively, cannot
be factorized, having
\beq
P_{n,n'}^{m,m'}(\delta )\neq P_n^m(\delta )P_{n'}^{m'}(\delta ),
\label{1bb}
\eeq
due to the existence of a 
bin-bin correlation between the bins $m$ and $m'$
\footnote{Strictly speaking, any statistical dependence 
between these bins can lead to property (\ref{1bb}).}.

A procedure for investigating such bin-bin correlations
is to measure so-called factorial correlators 
\cite{abir1,abir7,abir8,abir9},
(for a review see \cite{abir10}). In  terms
of  $P_{n,n'}^{m,m'}(\delta )$, $P_n^m$, and
$P_{n'}^{m'}$, the factorial correlators for two bins of equal size
$\delta$ can be written as
\beq
F_{q,q'}^{m,m'}(\delta )=
\frac{\sum_{n,n'}^{\infty}P_{n,n'}^{m,m'}(\delta )n^{[q]}\,{n'}^{[q']}}
{\left(\sum_{n=1}^{\infty}P_n^m(\delta )n^{[q]}\right)
\left(\sum_{n'=1}^{\infty}P_{n'}^{m'}(\delta ){n'}^{[q']}\right)},
\quad q',q>1,
\label{2bb}
\eeq
where $n^{[q]}=n(n-1)\ldots (n-q+1)$. The quantity in the numerator is
called the bivariate  factorial moment. In contrast to
the usual (univariate) factorial moment $\langle n^{[q]}\rangle =
\sum_{n=1}^{\infty}P_n^{m}(\delta )n^{[q]}$, which characterizes
only the local fluctuations in a single phase-space bin $m$, the
bivariate factorial moment contains  information on
correlation between the local 
fluctuations in the two bins, $m$ and $m'$.

If  no correlation exists between bins $m$ and $m'$,
we get $F_{q,q'}^{m,m'}(\delta )=1$ due to factorization  
of the multiplicity distribution in the numerator of (\ref{2bb}). 

To increase the statistics, one can assume translational 
invariance and average (\ref{2bb}) over all bin
combinations with the same bin-bin distance, $D$. After symmetrization,
one has 
\beq
F_{q,q'}(D)=\frac{1}{2(M-k)}\sum_{m=1}^{M-k}
\left(F_{q,q'}^{m,m+k}(\delta ) + F_{q',q}^{m,m+k}(\delta )\right),
\label{3bb}
\eeq
where $M=\Delta /\delta$, $\Delta$ is a full phase-space
interval, and $k=D/\delta$.

Correlators similar to (\ref{2bb}) have 
also been proposed in \cite{abir11}.
In this approach, the bin of size $\delta$ is divided into two parts.
If $n_{\mathrm{L}}$ and $n_{\mathrm{R}}$ are the number of particles
in the left part and the  right part of the bin, respectively, then 
one can define \cite{abir11}
\beq
F_2(M)=\frac{1}{M}\sum_{m=1}^{M}
\frac{\la n_{\mathrm{L}} n_{\mathrm{R}} \ra } 
{\la n_{\mathrm{L}}\ra \la n_{\mathrm{R}} \ra }.
\label{34bb}
\eeq

As is the case for the  usual univariate factorial moment, 
the multivariate factorial moments presented above are sensitive to the 
``tail'' of the multivariate multiplicity distribution
obtained in an experiment.
The limited statistics of an experiment 
reduce fluctuations measured by means of the 
high-order factorial moments  
because of the truncation of the multiplicity distribution 
\cite{abir12,abir13,abir14}.
This can 
exert a negative influence on the behavior of the factorial
correlators.

We note another shortcoming of the factorial correlators.
As the usual factorial moments, the multivariate
definition selects only ``spikes''. Dynamical information
from ``dips'', therefore, is completely lost. This means that
we lose important information on  bin-bin correlations.
As an example,  correlations should exist between
different bins that contain no particles, i.e., 
\beq
P_{0,0}^{m,m'}(\delta )\neq P_0^m(\delta )P_0^{m'}(\delta ).
\label{4bb}
\eeq
According to the definition, 
the factorial correlator is not able to measure  such
correlations.

The complete information on  bin-bin correlations can be obtained,
without
the bias arising from  restricted statistics of an
experiment, if one formulates
the problem in  terms of the bunching parameters 
\cite{abir3,abir4,abir5,abir6}.
The univariate bunching parameters for bin $m$ are defined in 
terms of the probabilities $P_n^m(\delta )$ as 
\beq
\eta_q^m(\delta )=\frac{q}{q-1}
\frac{P_q^m(\delta )P_{q-2}^m(\delta )}{(P_{q-1}^m(\delta ))^2}.
\label{41bb}
\eeq

Accordingly, it is possible to construct bivariate
bunching parameters in the same way as that done for bivariate
factorial moments,
\beq
\eta_{q,q'}^{m,m'}(\delta )=\frac{q q'}{(q-1)(q'-1)}
\frac{P_{q,q'}^{m,m'}(\delta )P_{(q-2),(q'-2)}^{m,m'}(\delta )}
{\left( P_{(q-1),(q'-1)}^{m,m'}(\delta )\right)^2},
\quad q,q'>1.
\label{5bb}
\eeq

The relation of BPs with usual 
moments have been found in \cite{abir3,abir5}.
For bivariate BPs, such a kind of relation can be written as
\beq
\eta_{q,q'}^{m,m'}(\delta )\simeq \frac{\la \rho_{m,m'}^{q,q'}\ra 
\la \rho_{m,m'}^{q-2,q'-2}\ra }{\la \rho_{m,m'}^{q-1,q'-1}\ra ^2},
\qquad \delta\to 0  
\label{54bb}
\eeq
due to the suppression of Poissonian noise in the limit of
small $\delta$.

As is the case for 
multi-dimensional probabilities, these quantities can be
expressed as
\beq
\eta_{q,q'}^{m,m'}(\delta )=\eta_q^m(\delta )\eta_{q'/q}^{m'}(\delta )=
\eta_{q'}^{m'}(\delta )\eta_{q/q'}^{m}(\delta ),
\label{55bb}
\eeq
where $\eta_q^m(\delta )$ is the usual univariate bunching parameter and
$\eta_{q'/q}^{m'}(\delta )$ represents a conditional bunching 
parameter for bin $m'$ 
constructed from  conditional probabilities, i.e., the
probability to observe $q'$ particles in bin $m'$ under the
condition that $q$ particles have been found in another bin $m$.
Then, the conditional BPs have the form
\beq
\eta_{q'/q}^{m'}(\delta )=\frac{q'}{(q'-1)}
\frac{P_{q'/q}^{m'}(\delta )P_{(q'-2)/(q-2)}^{m'}(\delta )}
{\left( P_{(q'-1)/(q-1)}^{m'}(\delta )\right)^2},
\quad q,q'>1.
\label{56bb}
\eeq

If the two bins are statistically independent, then the bivariate
bunching parameters factorize:
\beq
\eta_{q,q'}^{m,m'}(\delta )=\eta_{q}^m(\delta )\eta_{q'}^{m'}(\delta ).
\label{61bb}
\eeq
By analogy with the factorial correlators, the bunching correlators
can, therefore, be defined as
\beq
\breve{\eta}_{q,q'}^{m,m'}(\delta )=\frac{\eta_{q,q'}^{m,m'}(\delta )}
{\eta_q^m(\delta )\eta_{q'}^{m'}(\delta )}.
\label{6bb}
\eeq
As is the case for (\ref{2bb}), this 
definition  grants unity if the cells $m$ and $m'$ are
statistically independent.

The bunching correlators, in general, are not symmetric in $q$
and $q'$. As is performed in (\ref{3bb}), 
we can symmetrize this definition:
\beq
[\breve{\eta}_{q,q'}^{m,m'}(\delta )]_{\mathrm{S}}=\frac{1}{2}
(\breve{\eta}_{q,q'}^{m,m'}(\delta ) +
\breve{\eta}_{q',q}^{m,m'}(\delta )).
\label{7bb}
\eeq
Defining  the distance $D$ between  two bins, the bunching
correlators can further be averaged over 
many pairs of equidistant bins. In analogy to (\ref{3bb}),
the problem of bin-bin correlations can be formulated in terms
of the bunching correlators
\beq
\eta_{q,q'}(D)=\frac{1}{(M-k)}\sum_{m=1}^{M-k}
[\breve{\eta}_{q,q'}^{m,m+k}(\delta )]_{\mathrm{S}}
\label{8bb}
\eeq
and their behavior in the limit $D\to 0$.
 
According to the above definition of  bunching correlators,
the second-order bunching correlator contains 
important extra information {\em on empty bin-bin correlation}
that cannot be extracted 
by means of factorial correlators.
Indeed, if such correlations exist, then,
due to (\ref{4bb}), one obtains
\beq
\breve{\eta}_{q,q'}^{m,m'}(\delta )\neq 1
\label{9bb}
\eeq
for any combination such as $\{2,2\}$, $\{2,3\}$, $\{3,2\}$
etc. 
For the symmetrized and averaged bunching correlators,
this leads to
\beq
\eta_{q,q'}(D)\neq 1, \qquad q=2,
\quad q'=2,3\ldots .
\label{10bb}
\eeq
On the other hand, if only such (hypothetical) correlations exist,
the factorial correlators are equal to one for any higher
rank.  

\section{ The lowest-order bunching correlator  
and its behavior}
\label{sec:abi2}

The value of $\eta_{2,2'}(D)$ is affected by events  
having no particles in both bins and, hence, it incorporates the empty
bin-bin correlations that cannot be measured by means of factorial
correlators. In this section  we shall illustrate the 
dependence  of this quantity on the distance $D$ between the
two bins.

For our numerical calculations, we can rewrite the definition of 
$\eta_{2,2'}(D)$ as follows:
\beq
\eta_{2,2'}(D)=\frac{1}{M-k}\sum_{m=1}^{M-k}
\breve{\eta}_{2,2'}^{m,m+k}(\delta ),
\label{11bb}
\eeq

\beq
\breve{\eta}_{2,2'}^{m,m'}(\delta )=\frac{\eta_{2,2'}^{m,m'}(\delta )}
{\eta_2^m(\delta )\eta_{2'}^{m'}(\delta )}.
\label{12bb}
\eeq
To define bivariate and univariate BPs, 
we introduce the following expression
as an indicator for the presence of a given spike configuration
for a given experimental event $t$:
\beq
W_q(m, m',t)=
\left\{ \begin{array}{ll} 1, 
& \mbox {if both bins $m$ and $m'$ contain $q$ particles,} \\
0, & \mbox {otherwise.}
\end{array}
\right.
\label{13bb}
\eeq

Then, we have

\beq
\eta_{2}^{m}(\delta )=2
\frac{{\overline W}_2(m,m){\overline W}_0(m,m)}
{{\overline W}_1^2(m,m)},
\label{15bb}
\eeq

\beq
\eta_{2,2'}^{m,m'}(\delta )=4
\frac{{\overline W}_2(m,m'){\overline W}_0(m,m')}
{{\overline W}_1^2(m,m')},
\label{14bb}
\eeq

where ${\overline W}_q(m,m')$ is the average  
of $W_q(m, m',t)$ over $N_{\mathrm{ev}}$ experimental events
\beq
{\overline W}_q(m,m') =\frac{\sum_{t=1}^{N_{\mathrm{ev}}} W_q(m, m',t)}
{N_{\mathrm{ev}}}.
\label{16bb}
\eeq

An exact calculation of the statistical error (standard deviation)
is always a complex task and requires  special attention to 
any local measurement. Below, we give 
a sketch of propagation of 
the standard deviation for (\ref{11bb}).

The square of the 
standard deviation for ${\overline W}_q(m,m')$ is given by
\beq
S^2_q(m,m') =
\frac{1}{N_{\mathrm{ev}}(N_{\mathrm{ev}}-1)}
\left[\sum_{t=1}^{N_{\mathrm{ev}}}W_q^2(m, m',t) - 
N_{\mathrm{ev}}{\overline W}_q^2(m, m')\right].
\label{17bb}
\eeq
The square  of the standard deviation for second-order 
BPs is given by
\beq
V_2^2(m, m')=\frac{{\overline W}_{0}^2}
{{\overline W}_{1}^4}
s^2_2 + \frac{4{\overline W}_{2}^2{\overline W}_{0}^2}
{{\overline W}_{1}^6}s_1^2 +
\frac{{\overline W}_{2}^2}{{\overline W}_{1}^4}s_{0}^2.
\label{18bb}
\eeq
This expression gives us the square  of the standard deviation
for univariate BPs if 
\beq
{\overline W}_{q} = {\overline W}_{q}(m,m),\qquad
s^2_q = 4 S_q^2(m,m).
\label{19bb}
\eeq
The square  of the standard deviation
for bivariate  BPs can be found from  (\ref{18bb})
if
\beq
{\overline W}_{q} = {\overline W}_{q}(m,m'),\qquad
s^2_q = 16 S_q^2(m,m').
\label{191bb}
\eeq
The total statistical error for (\ref{11bb}) can be found
by combining the standard deviations for the univariate and
bivariate BPs and averaging the results over all bin pairs.

\begin{figure}
\begin{center}
\mbox{\epsfig{file=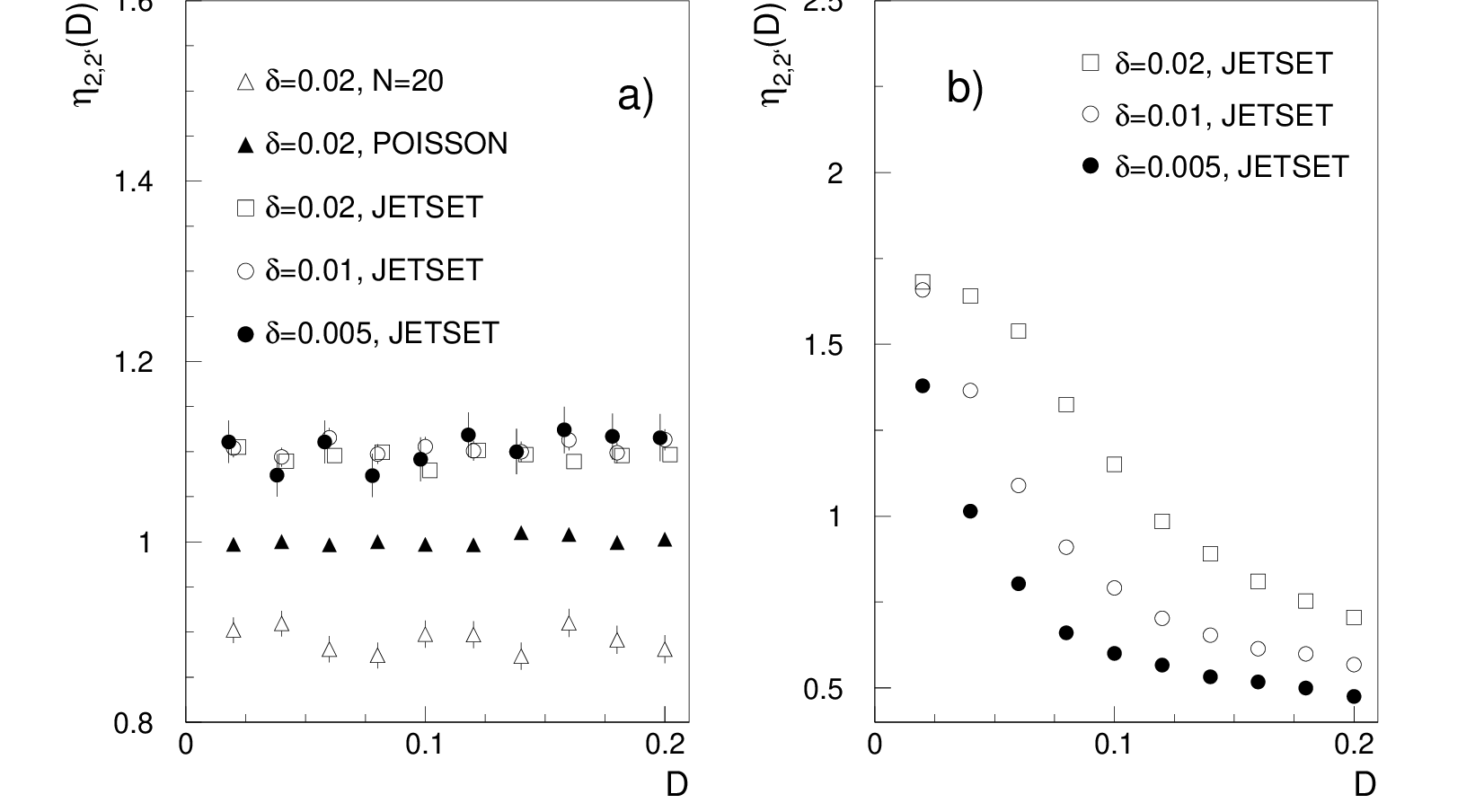,width=14.0cm}}
\caption[ap1]
{\it Value of $\eta_{2,2'}(D)$ as a function
of distance $D$ between bins. {\bf (a)} The behavior
in the case of purely statistical fluctuations for
different distributions of particles in full phase space.
{\bf (b)} The behavior for the case of
dynamical fluctuations (phase-space distribution
in azimuthal variable)
simulated by the JETSET 7.4 PS model.}
\label{ap1}
\end{center}
\end{figure}

In Fig.~\ref{ap1}a, the behavior of $\eta_{2,2'}(D)$ is shown
for the case of purely statistical phase-space fluctuations.
For our numerical calculations, 
we simulate the phase-space distribution
by a pseudo-random number generator in the
``phase space'' $0<x<1$. The total number of events
is 30,000. 
In this figure we consider the cases
in which a total number of particles $N$ in full phase space
fluctuates according to full-phase-space fluctuations.
We considered the following cases: 

1) $N$ is fixed for all events (N=20);

2) $N$ is distributed according to a Poissonian law 
   with mean ${\overline N}=20$;
   
3) $N$ is distributed according to the
   JETSET 7.4 PS model \cite{abir15} simulating 
   $\mathrm{e}^+\mathrm{e}^-$-annihilation at a c.m. 
   energy of 91.2 GeV. 
   Such a
   distribution is similar to a negative binomial.
   For this case, we also consider  different values
   of bin size $\delta$.

\medskip
   
As expected, the value of the 
bunching correlator is equal to 1 for the Poisson distribution.
We have 
verified  that this result is independent of
the mean of the Poisson distribution and of the bin size $\delta$.

For the sample with  fixed multiplicity ($N=20$), there is
a negative correlation, since $\eta_{2,2'}(D)<1$.
This kind of correlation is due to the trivial effect that
the probability of finding a particle in a bin is always less if
another particle has already been found in another bin. In the case
of no dynamical phase-space correlations, such a negative (pseudo) correlation
leads to a $D$-independent bunching correlator of value smaller than unity.

If particles are distributed according to a distribution
broader than Poisson, one should expect a positive
correlation. For the case of no phase-space correlations,
this again  leads to a $D$-independent bunching correlator,
but with a value of $\eta_{2,2'}(D)>1$.
       
In Fig.~\ref{ap1}b we present  
$\eta_{2,2'}(D)$ for a more realistic situation. 
Here,
$N$ again fluctuates according to JETSET 7.4 PS, but the 
phase-space distribution is defined in
the azimuthal angle with respect to the 
$\mathrm{e}^+\mathrm{e}^-$ collision axis.
To compare the results with the previous cases, 
this variable (with full phase-space range $2\pi$) 
has been transformed to a new variable with
unit range.  
Due to the jet structure of single 
events, the phase-space distribution
in this variable 
contains  dynamical fluctuations. 
As can be seen from Fig.~\ref{ap1}b, such
fluctuations lead to a bin-bin correlation. The correlation
increases for decreasing distance $D$, 
from $\eta_{2,2'}(D)<1$ for large $D$ to
$\eta_{2,2'}(D)>1$ for small $D$.
Moreover, in contrast to Fig.~\ref{ap1}a, the value of 
$\eta_{2,2'}(D)$ is affected by
the value of the bin size $\delta$.

\section{Conclusions}

In this paper, the bunching-parameter method has been
extended to measure bin-bin correlations. This
application of the bunching-parameter method 
has been achieved by considering bunching correlators
in analogy to factorial correlators.
The method not only allows one  to study 
fluctuations inside a phase-space bin without experimental
bias from finite statistics, but also to study correlations between
bins separated in phase-space.

One of the remarkable features of the bin-bin correlation
study is that the main properties of local fluctuations
inside bins, and correlations between the bins can be formulated
in a unified manner. Based on our analysis of second-order
bunching correlations and on \cite{abir5}, we conclude:

1) For purely statistical phase-space fluctuations,
the values of the univariate bunching 
parameters and those of the bunching
correlators are independent of bin size and bin-bin distance.  
These values
are affected by event-to-event multiplicity fluctuations, but are
equal to unity for Poisson-distributed particle multiplicity in full
phase space;

2) For dynamical phase-space fluctuations, the values of
univariate bunching parameters, and bunching
correlators increase for decreasing bin size $\delta$ or  
distance $D$ between two bins.

Such a similarity in the behavior of these quantities is the result of
an intrinsic relation between fluctuation and correlation
properties of the local fluctuations.  

Finally, from our study, let us note that  no
universal scaling relation between the local fluctuations and
correlations is observed for the azimuthal-angle distribution in
JETSET 7.4 PS model,
as it follows from the random-cascade model \cite{abir1,abir2},
for which the factorial correlators are $\delta$-independent.
The analysis of bin-bin correlations based on 
the bunching correlators
clearly shows that the behavior of the second-order correlator
is affected by the bin size $\delta$. In fact, this means that 
realistic intermittent fluctuations cannot be fully
described by the scaling indices of the univariate normalized 
moments as  is
the case for the random-cascade model.
For this reason, 
the experimental measurement of the correlators is an important 
complementary  part  of  the fluctuation analysis,  which, therefore,
cannot be reduced to the investigation  of the scaling indices of the
local quantities only.

\bigskip

Acknowledgments

\medskip

This work is part of the research program of the ``Stichting voor
Fundamenteel Onderzoek der Materie (FOM)'', which
is financially supported by the ``Nederlandse Organisatie voor
Wetenschappelijk Onderzoek (NWO)''.
We thank W.Metzger for useful remarks.

{}

\end{document}